\documentclass[aps,prb,preprint]{revtex4} 
\usepackage{graphicx}
\begin{document}
\title{A molecular perspective of water at metal interfaces}
\author{Javier Carrasco$^1$, Andrew Hodgson$^2$, Angelos Michaelides$^3$}
\affiliation{
$^1$Instituto de Cat\'alisis y Petroleoqu\'imica, CSIC, Marie Curie 2, E-28049, Madrid, Spain \\
$^2$Surface Science Research Centre and Department of Chemistry, University of Liverpool, Oxford Street, Liverpool, L69 3BX, UK \\
$^3$Thomas Young Centre, London Centre for Nanotechnology and Department of Chemistry, University College London, London WC1E 6BT, UK }

\begin{abstract}

Water-solid interfaces are ubiquitous and of the utmost importance to industry,
technology and many aspects of daily life.
Despite countless studies from different areas
of science, detailed molecular-level understanding of water-solid interfaces
comes mainly from well-defined studies on flat metal surfaces. 
These studies have recently shown that a remarkably rich variety of structures
form at the interface between water and seemingly simple flat metal
surfaces.
Here we discuss some of the most exciting examples of recent work in this
area and the underlying physical insight and general concepts that emerge 
about how water binds to surfaces.
A perspective on the outstanding problems, challenges, and open questions in the
field is also provided.

Copyright $\copyright$ 2012, Rights Managed by Nature Publishing Group
\end{abstract}

%\pacs{
%68.43.Bc,        % Ab initio calculations of adsorbate structure and reactions
%68.43.Fg,        % Adsorbate structure (binding sites, geometry)
%71.15.Mb 	 % Density-functional theory, LDA, GGA and other corrections
%}

\maketitle

\section{Introduction}

Water-solid interfaces are important to an incredibly broad range of everyday phenomena
and scientific and technological processes. 
This includes areas such as corrosion, 
electrochemistry, nanoparticle self-assembly, environmental chemistry, lubricants, and 
heterogeneous catalysis, to name just a few.
Indeed the ubiquitous
presence of water on surfaces under ambient conditions means that water-solid interfaces
are relevant to many areas of the physical sciences. 
In addition, contemporary issues
such as climate change and shortages in both water and energy mean that there is
a pressing need to better understand the structure and dynamics of water-solid interfaces.
This, along with fundamental questions about hydrogen (H) bonding and H atom transfer\cite{michaelides2007,kumagai2012}, 
has resulted in a flurry of interest in recent years and the chemistry and physics of
interfacial water is now one of the most exciting and thriving fields in materials chemistry
and physics.

The first step towards understanding water-solid interfaces is the challenging task of
 knowing where the atoms and molecules are located. 
 This is challenging because of the extremely small
 number of atoms at the interface (relative to the bulk) and the difficulty of analyzing the
 details of atomic-level geometries and electronic structure for such a small sample
 \cite{campbell2011}. 
 Typically this requires what is generally referred to as a surface science style approach, i.e., 
 extremely clean, atomically smooth substrates are 
 interrogated under ultra-high vacuum (UHV) conditions to determine the arrangement of 
 water and how it bonds to the surface. 
 Of course we do not live under
 UHV conditions, but if we want to obtain precise atomic-level understanding and rigorously
 ``solve'' water-surface structures, UHV provides the armoury of tools we need. 
A disadvantage of this approach is that water is only stable  at cryogenic
 temperatures (up to about 200 K) in UHV, so that the structures which form are frozen, without any of the 
 dynamic disorder of a liquid. A second experimental limitation is that often only the first contact layer is examined, 
 with few experiments able to probe thicker water layers in any detail. 
 A consequence of this is that understanding how the interface structure responds to multilayer 
 adsorption, or to melting, requires us to develop accurate, predictive theoretical models for the solid water 
 interface based on the few systems studied in detail.
 Thus, although there is broad interest in water-solid interfaces from many disparate areas of 
 science, the fundamental atomic-level insight into water-solid interfaces is currently limited to a small number
 of surface science style studies, some of which we discuss in this article.

 Probably the biggest conceptual idea
 to emerge from well-defined studies of water at interfaces is that of 
 the \emph{bilayer adsorption model}. 
 The bilayer model 
is a concept that has been used to discuss interfacial water in almost all environments and is 
 often the first concept that newcomers to the field grasp. Perhaps as a consequence, the 
  bilayer model has gained a currency that is entirely out of line with the experimental evidence in its 
  favour, often being the last concept to be abandoned only
 in the face of overwhelming evidence! 
 The term bilayer is used both in its strict sense
 --- a buckled hexagonal overlayer resembling the basal plane of ice --- and more loosely
 to discuss structuring of water films at interfaces where two peaks in the density profile
 are observed.
 %
 % AM took this out. I didn't really understand. reminiscent of a liquid. 
 %
 The bilayer model was originally developed based on experiments on water at metal surfaces and this article will explore 
 some of the groundbreaking new work published in this area recently 
 which calls this model for water adsorption into question. 
 Indeed, here we will consider
 if a water bilayer in the strict sense of the word has ever been observed, or should really be expected.

 In the following we will explore what the bilayer model is, how it
 emerged, and how recent experiments and calculations reveal a much more interesting
 variety and richness of structure for water at interfaces. 
 The focus will be on water--metal interfaces, the system where the first structural models for water 
 at interfaces emerged and where some of the most detailed and exciting recent experiments and 
 simulations have been performed. 
 Key recent developments include the observation of wetting layer structures built exclusively from
 pentagons, or from combinations of heptagons and pentagons, the observation
 of wetting layer structures rich in defects in the H bonded network, 
 the recognition of the importance of strain in the H bonded network,
 and the realisation that maximising the number of H bonds does not
 necessarily lead to the most stable overlayer structures. 
 The challenges and issues that remain before a fully predictive understanding
 of the structure and dynamics of water-solid interfaces can be achieved will also be discussed.
 This will be a short
 review aimed mainly at newcomers to the field, using a few selected examples to illustrate the concepts and exciting 
 developments alluded to above. 
 For comprehensive overviews of the literature up to 2009 the interested reader is referred to
 refs. [\cite{Henderson2002,hodgson2009}] and for other complementary mini-reviews on some 
 specific aspects of water at surfaces to refs. [\cite{feibelman2009,feibelman2010,michaelides2006}].

 \section{A brief history of water on metals: the rise of the bilayer}
 
\begin{figure}[htb]
\includegraphics[width=0.90 \textwidth]{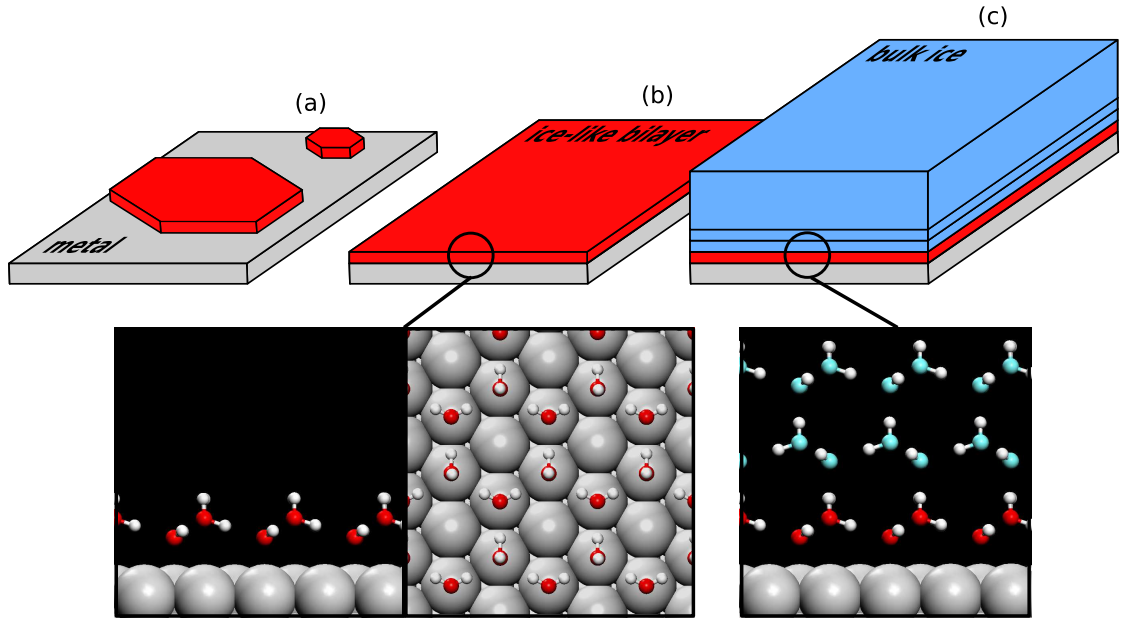}
\caption[]{\textbf{Schematic (upper frames) and molecular models of the traditional model for water/ice adsorption on a solid surface.} 
(a) Water is assumed to form extended 2D islands of the ice bilayer structure that (b) forms a wetting layer in registry 
with the close packed surface. Ice multilayers then grow on top of the first layer, forming a strained epitaxial arrangement (c).}
\label{fig:bilayer_model}
\end{figure}

 As we have said, obtaining atomic-level understanding of surface structure is very 
 challenging. 
 For water-solid interfaces this generally involves preparing clean atomically flat
 surfaces upon which water is deposited under UHV conditions. 
 This frozen in
 structure --- typically a single layer --- can then be interrogated with, for example, 
 techniques such as low energy electron diffraction (LEED). 
 In the 1980s Doering and 
 Madey performed such a study \cite{doering1982}, examining in meticulous detail water
 adsorbed on the close-packed (0001) surface of Ru. This surface has an hexagonal
 arrangement of metal atoms with the distance between neighboring Ru atoms 
 within 3 \% of the nearest neighbour water-water distance in ice I. 
 Upon observing a $\sqrt3\times\sqrt3-R30^\circ$ LEED 
 pattern the bilayer model was introduced for extended 2D overlayer adsorption
 of water on metals. 
 The bilayer model proposed by Doering
 and Madey \cite{doering1982} consists of an epitaxial 2D arrangement of water 
 molecules forming an {\it hexagonal} network, which resembles the (001) basal plane
 of ice I (see Fig. \ref{fig:bilayer_model}).
 It is referred to as a bilayer simply because the hexagonal rings within the (001)
 basal plane of ice are puckered with molecules at two distinct heights. 
 At a metal surface it was suggested that the water molecules within the lower part of the bilayer 
 interact relatively strongly with the surface, whereas the water molecules in 
 the upper part of the bilayer (about 1 \AA\ higher) are H bonded to the bottom layer
 and do not interact appreciably with the surface. 
 In a conventional bilayer the molecules in the top half of the bilayer have OH bonds
 which point away from the surface, often referred to as dangling OH groups or the H-up bilayer model. 
 An alternative arrangement in which the OH groups are directed at the surface (``H-down'')
 has also been suggested \cite{ogasawara2002,hodgson2009}. 
 In either case it's interesting to note that these are so called proton ordered structures
 when built within a two molecule $\sqrt3$ unit cell and as such are bilayers of ice XI, the 
 proton ordered form of ice I. 
 
 Following Doering and Madey, $\sqrt3$ LEED patterns observed on other metal
 surfaces \cite{hodgson2009} were interpreted as evidence of the generality of the bilayer model. 
 In  addition an apparent correlation between lattice mismatch and
 binding energy --- obtained by monitoring desorption temperatures of water from
 various metals --- was taken as further evidence for an epitaxial bilayer in which the
 intermolecular water-water distances adapted to the lattice constant of the underlying substrate
 \cite{thiel1987}. 
 The bilayer model was also adapted to explain the (2$\times$2) diffraction patterns observed 
 on  f.c.c. (110) surfaces \cite{hodgson2009}. 
 Essentially the bilayer model became the standard model for water adsorption and was
 subsequently used to interpret 
 surface structures on all sorts of inorganic substrates.
 It is still widely used.
 
  \section{The first signs of a problem...}
 
 In the 1990s Held and Menzel reported a more sophisticated LEED
 measurement that led to the first structural characterization of a water overlayer on any surface \cite{held1994}.
 Using a newly developed low current LEED apparatus they performed what's known as a LEED-IV study. 
 By comparing LEED-IV spectra
 with simulated spectra obtained from trial structural models, a 3D structural model of the 
 interface can be arrived at describing
 the location of the heavy atoms at the interface. (Because of the light
 mass of hydrogen their positions are not readily determined in LEED \cite{held2001}, another reason why 
 in general solving structures for water is difficult). 
 Held and Menzel found
 that the best fit to their data was obtained when the oxygen atoms were almost
 co-planar. 
 This non-buckled ``monolayer'' is the antithesis of a bilayer wherein the O atoms are buckled
 by almost 1 \AA\  to gain the tetrahedral coordination found in ice I. 
 Thus the very first structural characterisation 
 of water at an atomically flat interface conducted almost 20 years ago was actually inconsistent
 with the bilayer model.

One of the first people to think about this apparent contradiction was Feibelman \cite{feibelman2002}
who in 2002 reported density functional theory (DFT) calculations for water on Ru. 
He suggested that the small buckling was a sign that some of the water
 molecules in the overlayer had dissociated yielding a so called partially dissociated 
 overlayer, similar to an overlayer reported earlier for water on Pt(111) \cite{michaelides2001,michaelides2001b}.
Feibelman's calculations indicated that partially dissociated structures
containing both OH and H$_2$O are significantly more stable than any conventional 
intact bilayer picture and suggested that what Held and Menzel had 
observed was a partially dissociated overlayer. 
This suggestion prompted 
a flurry of interest and water on Ru quickly became the hottest system 
for investigation in the 2000s.
The reader interested in the details
of this controversial and at times heated discussion is referred to Hodgson's
review \cite{hodgson2009}. 
For the present discussion it is sufficient to say 
that, somewhat ironically, the $\sqrt3$ 
diffraction pattern responsible for the birth of the bilayer model does not actually result 
from a conventional bilayer \cite{gallagher2009}. 

 \section{The STM revolution: the fall of the bilayer}
 
\begin{figure}[htb]
\includegraphics[width=0.80 \textwidth]{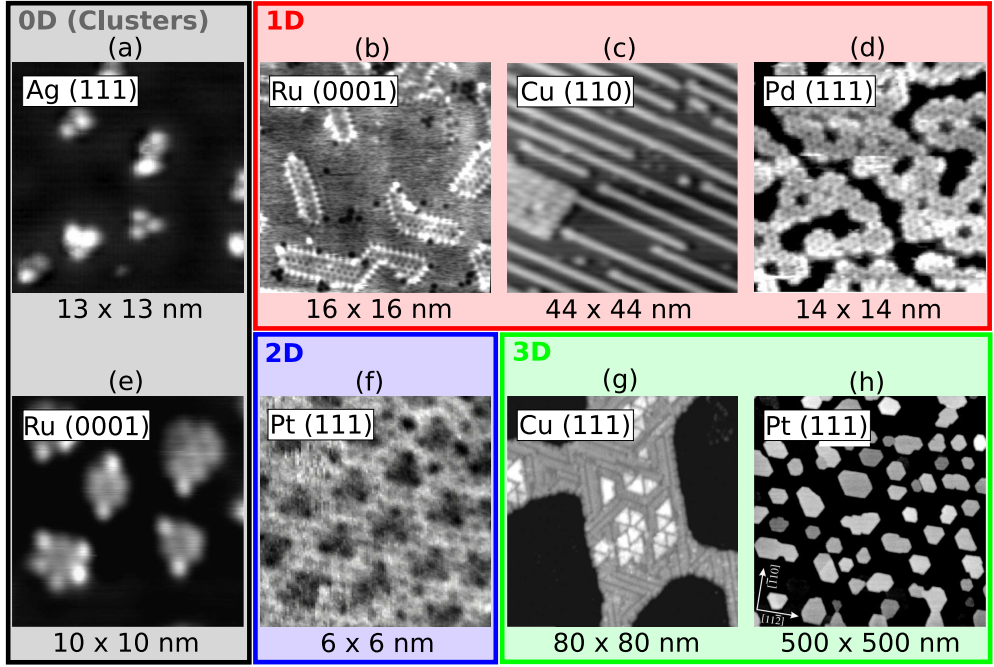}
\caption[]{\textbf{Experimental STM images of water clusters and overlayers on metals.} A selection of structures with different
dimensionality are shown: (a,e) clusters of about 6 to 20 water molecules on Ag(111) and Ru(0001); (b,c,d) strips,  long chains,
and rosette meshes, (f) 2D network, and (g,h) crystalline 3D pillars and pyramids. Adapted from
Gawronski {\it et al.} \cite{gawronski2008} (a),
Tatarkhanov {\it et al.} \cite{tatarkhanov2008} (b),
Yamada {\it et al.} \cite{yamada2006} (c),
Cerda {\it et al.} \cite{cerda2004} (d),
Salmeron {\it et al.} \cite{salmeron2009} (e),
% Sachs {\it et al.} \cite{sachs2002} (f),
Nie {\it et al.} \cite{nie2010} (f),
Mehlhorn {\it et al.} \cite{mehlhorn2007} (g),
and Th\"urmer and Bartelt \cite{thurmer2008} (h).
\label{fig:stm_structure}}
\end{figure}

 With the development of new techniques --- most notably scanning 
 tunnelling microscopy (STM) --- it became possible to simply ``see'' at a local level what structures
 water forms when it sticks to metal surfaces. 
 Such experiments, when interpreted together with DFT (and sometimes also with spectroscopic or diffraction measurements), have 
 transformed our understanding of water-solid interfaces. 
 A selection of recent low temperature STM results for water on metals is shown in Fig. \ref{fig:stm_structure}.
 One of the most interesting points to note from all these experiments is the tremendous diversity of
 structures observed. 
 Although the experiments were all performed at low temperatures on atomically flat
 metal surfaces, on each surface a  different structure is observed.
 On some surfaces (generally at lower coverages) isolated water clusters are observed. 
 On others, 1D or quasi-1D chains are observed, on others 2D overlayers with mixtures of pentagons 
 and heptagons, and on others 3D pyramids and towers. 
 In no case is a uniform 2-D ice bilayer detected. 
We now discuss some of the systems shown in Fig. \ref{fig:stm_structure} and the concepts 
and physical understanding that have emerged. 

\subsection{Water clusters: ``Zero-dimensional'' structures}

\begin{figure}[htb]
\includegraphics[width=0.60 \textwidth]{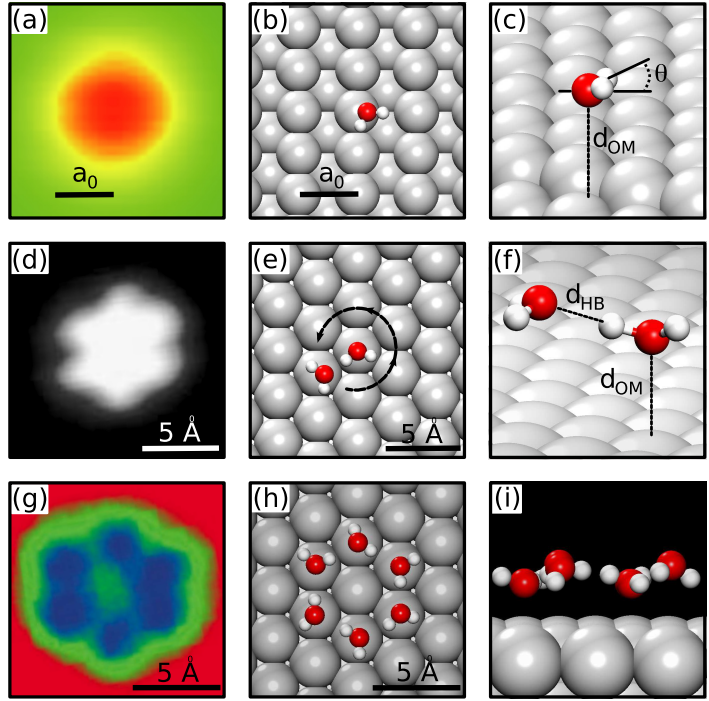}
\caption[]{\textbf{Experimental STM images and geometrical models predicted by theory for a water monomer and water clusters on
various metal surfaces.}
The experimental images correspond to a water monomer on Cu(110) (a), a water dimer on Pt(111) (d), and a water hexamer on Cu(111) (g).
Top and side views of the corresponding theoretical structures are shown in the central and right hand columns. 
It has been suggested that the flower-like shape in the experimental STM image of the water dimer (d) is a consequence of 
rapid helicopter-like rotation of the water dimer about the plane of the surface normal as indicated in (e).\cite{motobayashi2008} 
Scale bars are indicated on each of the images where useful, either reporting distances in Angstrom ((d), (e), (g), (h)) or in terms 
of the underlying lattice constant (a$_0$, panels (a) and (b)).
Adapted from Okuyama {\it et al.} \cite{okuyama2011} (a),
Motobayashi {\it et al.} \cite{motobayashi2008} (d), and 
Michaelides and Morgenstern \cite{michaelides2007} (g).
\label{fig:clusters}}
\end{figure}

% Individual water molecules were first imaged on a metal surface with STM on Pd(111)\cite{mitsui2002} (Fig. \ref{fig:clusters}(a)). 
%  AH Don't like this paper being used like this, its very unlikely they ever saw a monomer, they're at 40 K, 
% and who knows what the "dimer, trimer and tetramer" are! 
Individual water molecules are extremely difficult to image on metal surfaces due to their high mobility and
preference to form H bonded clusters. Water monomers can be identified at very low coverage and temperatures (below 20 K) \cite{motobayashi2008,kumagai2009,kumagai2012}, sitting in the atop sites ($i.e.$, the sites directly above individual metal atoms of the substrate). 
The atop sites were also predicted to be the most stable adsorption sites with density functional theory (DFT) 
on both close-packed and more open metal surfaces \cite{michaelides2003,ranea2004,meng2004}.
At this adsorption site, DFT further predicts that the dipole moment of the water molecule is 
aligned almost parallel to the surface (Fig. \ref{fig:clusters}(c)). 
This binding mode favours interaction of the 1b$_1$ molecular orbital of water with the surface \cite{michaelides2003,carrasco2009b}.

% AH Again, be better if the image 3a was demonstrably of a water monomer, eg from Motobayashi SS, Kumagai PRB or Okuyama JPD?

Even at these low temperatures, however, water diffuses rapidly across metal surfaces and readily
forms H bonded clusters.
The simplest cluster, the water dimer, has been observed on a number of metal 
surfaces \cite{kumagai2008,motobayashi2008} 
and DFT studies suggest that it has an asymmetric structure with the two 
water molecules at different heights above the surface (Fig. \ref{fig:clusters}(f)).
The water molecule donating the H bond is more tightly bound and closer to the surface (generally by about 0.5 \AA)
than the water molecule accepting the H bond. 
It has been suggested that this asymmetric structure facilitates facile
H bond rearrangement, rotation, and diffusion of water dimers \cite{ranea2004,kumagai2008,motobayashi2008,michaelides2006}.

Adsorbed water trimers and tetramers
% and pentamers
have all also been observed \cite{kumagai2011,okuyama2011}, with the tetramer being 
the first cyclic cluster that is more stable than a simple
linear chain on Cu(110). 
However, the 
cluster which has received the most attention is the cyclic water hexamer.
Interest in hexamers stems partly from the fact that they are seen as the building blocks of ice  
and, in the past few years, hexamers have
been successfully resolved with
STM (often in conjunction with complementary DFT studies) on Pd, Cu, and Ag surfaces
\cite{mitsui2002,morgenstern2002,morgenstern2002b,michaelides2007,gawronski2008,mehlhorn2009}. 
Although in the gas phase water hexamers possess a number of almost iso-energetic isomeric 
forms (notably the ``cage'', ``book'', ``prism'', and ``cyclic''' isomers \cite{biswajit2008}), when observed 
on metal surfaces only cyclic water hexamers have been detected to date.
In these hexamers each water molecule is located near the favoured atop adsorption 
site and accepts and donates just a single H bond.
Adsorbed cyclic hexamers, therefore, have a different H bonding topology
to a hexamer extracted from a bilayer of ice ($c.f.$ Figs. \ref{fig:bilayer_model} and \ref{fig:clusters}(h)).

Depending on the substrate, the hexamers are either planar (with all molecules at the same height) or 
buckled (with molecules at two distinct heights, as shown in Fig. \ref{fig:clusters}(i)). 
Generally planar hexamers are favoured on reactive surfaces to which water molecules
bond relatively strongly (e.g. Ru) \cite{haq2006}, whereas on the noble metals (Cu and Ag) \cite{michaelides2007} 
the buckled hexamer 
is favoured. 
This difference points to the subtle balance between water-water and 
water-metal interactions in directing the structure of these adsorption systems. 
Indeed, as we will see, this balance is the single most important issue in determining the
structure of water clusters and overlayers on metals. 
In the DFT calculations of water hexamers on Cu and Ag the balance between water-water and 
water-metal interactions was examined in detail and it was shown that there
was a competition between the ability of water molecules to
simultaneously bond to a substrate and to accept H bonds. 
This comes about because the same molecular orbital involved in bonding to the surface (1b$_1$) is
also implicated in the acceptance of H bonds. 
This competition was in turn used to rationalize previous structure predictions
for other water clusters adsorbed on various surfaces, including the asymmetric buckled 
structure of the water dimers and higher clusters \cite{michaelides2007}. Loosely speaking, we can think of planar 
hexamers as optimizing the interaction between the water and the metal on reactive surfaces, whereas  
buckled structures optimize the H bonding within the cluster on noble metals. As we will see shortly, the formation 
of planar networks on reactive surfaces places constraints on how large a water cluster can grow before the 
structure is forced to buckle out of plane, reducing the bonding to the surface.

\subsection{One-dimensional structures}

\begin{figure}[htb]
\includegraphics[width=0.80 \textwidth]{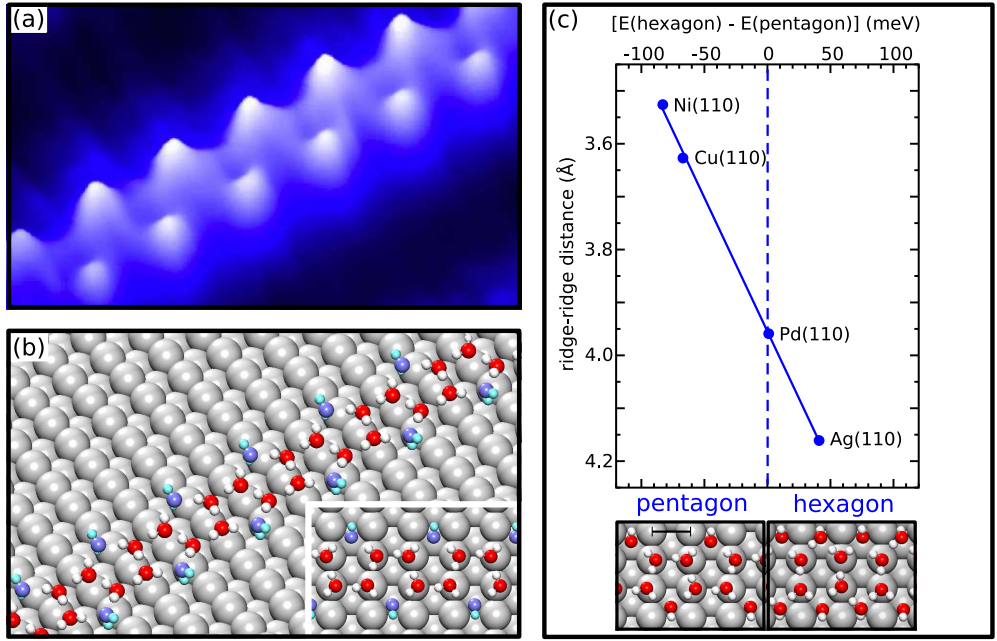}
\caption[]{\textbf{STM and DFT results for water on Cu(110).} (a) STM image of the $ca.$ 1 nm wide ice chains that
nucleate on Cu(110). (b) Structural model derived for the chains, which comprises a zig-zag
arrangement of water pentagons. It is the blue water molecules in (b) that give rise to the brightest features in (a). 
(c) Illustration of how the relative stability (in meV/H$_2$O) of hexagon and pentagon based 1D 
chains depends on the lattice constant of the underlying substrate (expressed in terms of the distance between the
close-packed 110 ridges. This shows that according to DFT 
pentagons are favoured on substrates with a 
relatively small lattice constant (Ni and Cu), whereas hexagons are favoured on the substrate with the
largest lattice constant, Ag. Figure based on results reported in ref. \cite{carrasco2009}.}
\label{fig:pentagons}
\end{figure}

We move our attention now to structures with periodicity in one dimension. 
To this end we focus on recent studies of  $ca.$ 1 nm wide ice chains that 
nucleate on Cu(110)\cite{yamada2006,lee2008,carrasco2009}.  
A particularly striking example of one of the chains --- which form spontaneously when water is 
adsorbed on Cu(110) at submonolayer coverages --- is shown in Fig. \ref{fig:pentagons}. 
Early suggestions for the structure of the chains were based on a bilayer-like arrangement of 
water molecules or other hexagonal structures\cite{yamada2006}.
%,schiros2008}. 
%
However, it was subsequently shown, through a combination of STM, 
infrared spectroscopy and DFT, that
the chains are not built from hexagons, but instead are built from a face-sharing
arrangement of water pentagons \cite{carrasco2009}.
Although there is a precedent for pentagon-based ice structures
in other environments \cite{ma2004,naskar2005}, this was the first time that such a
structural unit had been seen at a surface. 
The nucleation of
pentagons revealed an unanticipated structural adaptability of water--ice films at surfaces,
which goes well beyond the simple bilayer model.

The pentagon structure is favoured over others (e.g. the traditional hexamer building block) 
because it maximises the water--metal
bonding while maintaining a strong H bonding network. 
Specifically, in the pentagon chains two thirds of the water molecules bond at
atop adsorption sites of the Cu ridges, a higher proportion of 
molecules at the preferred adsorption sites than found in other structures. 
In addition, the pentagon structure allows for the strongest H bonded network 
with minimal strain within the overlayer. 
Indeed, fitting a commensurate network of
hexagons across the 110 rows of the Cu substrate as chains would require a $c.a.$ 1 \AA\ 
compression of each hexagon compared to the ideal hexagonal geometry of
ice. This introduces strain in the overlayer, which thus destabilizes
structures based on hexagons.
Put simply, the slighter smaller pentagons make a better fit to the 
Cu substrate than do hexagons.

With the above in mind DFT was also used to predict the structure and relative 
stability of 1D chains of hexagons and pentagons on the (110)
surfaces of Ni, Pd and Ag. 
Along with Cu, this consists of a set of
four (110) substrates with a separation between the close-packed
110 ridges that ranges from about 3.5 \AA\ to 4.2 \AA\ . As
shown in Fig. \ref{fig:pentagons}(c) an interesting correlation between the relative
stability of hexagon versus pentagon chains was observed. 
Pentagons are favoured on substrates with small lattice constants; that is, in
addition to Cu(110), Ni(110) is identified as a substrate on which pentagons
may nucleate. Pd(110) is a borderline case, whereas on the substrate
with the largest lattice constant Ag(110), there is a small preference for
hexagons.
This trend holds for the same water structures in 
the gas phase (that is, without the presence of the substrate) \cite{carrasco2009},
which provides further support that H bonding is key to the relative stability of
hexagons versus pentagons.

The water networks that form when about 0.5 ML of water is dosed on the close-packed (111) of Pd at low temperature are an intriguing
example of a structure that is intermediate between 1D and 2D,
as shown in Fig. \ref{fig:stm_structure}(d). 
The water/Pd system has already been reviewed
\cite{verdaguer2006,michaelides2006,hodgson2009,feibelman2010}
and so we point out just the key features of this system here. 
Essentially this structure is mainly comprised of long chains of hexagonal units,
whose width never exceeds a few hexagonal cells, as shown in Fig. \ref{fig:stm_structure}(d). 
When one uses DFT to work out the structure of the chains it 
turns out that the
overlayer is again not based on the puckered hexagonal structure of ice I.
Instead, it involves an arrangement of water molecules wherein most of the molecules are
at their preferred atop adsorption sites in their preferred orientation almost parallel 
to the surface.
Having water molecules flat against the surface places too many O-H bonds 
in the plane parallel to the surface and so a structure dominated by flat lying water molecules is 
only possible in the submonolayer regime with, in this case, a highly defective, metastable H bonded 
network. Completing the network forms a complex, partially ordered 2D structure made up of flat and 
H-down water \cite{mcbride2012}.

\subsection{Two-dimensional structures}

\begin{figure}[htb]
\includegraphics[width=0.60 \textwidth]{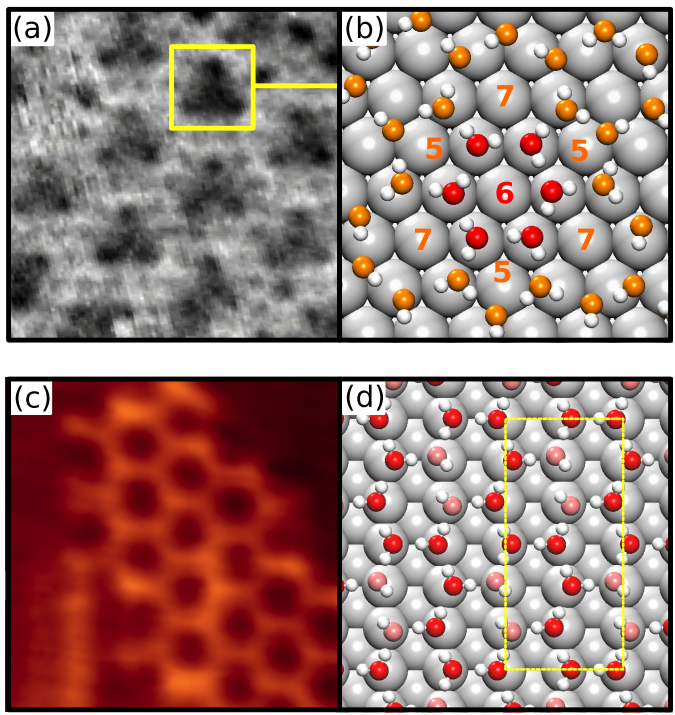}
\caption{\textbf{STM and DFT results for extended water overlayers on two different metal surfaces.} STM image (a)
of a water adlayer on Pt(111) and the corresponding DFT-based structural model (b).  
The structural model is comprised of 5, 6, and 7-membered rings with the water molecules at various 
heights above the surface (from ref. \cite{nie2010}).
STM image (c) of a water-hydroxyl overlayer on Cu(110) and a DFT-based 
structural model (d) for this overlayer (from ref. \cite{forster2011}). 
The periodic overlayer model is comprised of a 2:1 H$_2$O-OH ratio
with a high concentration of Bjerrum D defects (situations where 2 Hs sit between adjacent O atoms).}
\label{fig:overlayers}
\end{figure}

The 2D water monolayer that covers Pt(111) at low temperature
\cite{glebov1997} is the perfect example of an extended overlayer which clearly
deviates from the ice-like bilayer picture. 
After a long standing effort, in 2010 two
independent studies were able to obtain high resolution STM images of this water
layer \cite{nie2010,standop2010} (see Fig. \ref{fig:overlayers}(a)).
The current interpretation of this data, 
essentially supported by DFT calculations, seems to indicate that the 
wetting layer on Pt(111) is a mixture of pentagonal, hexagonal, and heptagonal
rings of water molecules \cite{nie2010,feibelman2010b} (Fig. \ref{fig:overlayers}(b)).
Similar to the isolated hexamers and 1D chains in previous sections, the balance
between water-metal and water-water interactions drives the H bonding 
structure adopted. The unit cell contains ca. 26 H$_2$O molecules, where just 6 are lying flat on 
atop Pt sites; the rest are bonded well above the surface and are tilted out of the surface plane. This layer has a 
substantial fraction of dangling H bonds, mostly pointing down towards the surface, 
and preserves a complete H bonding network.
Overall this configuration appears to be favoured by allowing a small fraction of the water molecules to bind flat, close to the 
metal substrate, maximizing their water-metal interaction while still completing a perfect H bonded network of water molecules.

The wetting of Cu(110) offers another recent example where the ice-like bilayer 
model does not hold. Above 140 K experiments indicate the formation of a
partially dissociated water layer, that is ordered yet not
fully stoichiometric (with ca. 2:1 H$_2$O-OH ratio) \cite{hodgson2009}. Although 
ice-like bilayer models have been used to attempt to explain this structure
in the past \cite{ren2006}, recent insights reveal a more complex scenario
\cite{forster2011}. Interestingly, the overlayer is not composed of a fully H bonded
network, instead it contains Bjerrum defects, where two H atoms sit between adjacent O atoms. 
This arrangement implies that maximizing the number of H bonds 
within a wetting overlayer is not a reliable criterion to identify stable structures. In
this particular case, the overlayer is stabilized by the formation of strong
H bonds as water donates to hydroxyl, with pairs of non-donor hydroxyls 
accommodated as defects within the H bond network (Fig. \ref{fig:overlayers}(d)).
This explains the
formation of a non-stoichiometric overlayer preferentially containing 
(H$_2$O)$_2$OH trimers and giving rise to the 2:1 H$_2$O-OH ratio. 
Hydroxyl groups are 
strong H bond acceptors, but rather weak donors, a principle that also helps to explain the 
1:1 H$_2$O-OH chains that form in the absence of excess water 
\cite{forster2012}, as well as the structure of hydroxyl/water phases on other surfaces \cite{schiros2010}.

\subsection{Three-dimensional structures}

On non-wetting surfaces such as Cu(111) water forms 3D clusters, which show a variety
of complex structures, but never a simple Ih(0001) or Ic(111) bilayer termination \cite{mehlhorn2007}. 
Depending on coverage, the structures formed include monomer decorated double 
bilayers with different super-structures, a faceted surface, pyramidal islands, and 
nanocrystallites with well defined height. Instead of the simple bilayer termination, 
a (2$\times$1) superstructure and pyramidal facets of Ih(1101) or Ic(221) are the most commonly observed terminating motifs.
A superstructure was also seen in He atom scattering from ice layers 
 grown on Pt(111) \cite{glebov2000} and has been explained by the ordering of dangling 
 OH groups in the ice surface to minimise the surface energy \cite{buch2008,pan2008,pan2010}. 
 
 Metal surfaces that do wet show a range of multilayer adsorption behaviour, depending on 
 how tightly the first layer of water is bound. On several surfaces where stable wetting 
 layers form, such as Ru(0001) and Pd(111), adsorption of further water causes 3D ice 
 clusters to grow, rather than a continuous water film \cite{Haq2007fr, kimmel2007}. 
 These surfaces are characterised by having tightly bound first layer water, with no 
 dangling OH groups to bond to further water. Evidently it is not favourable to 
 disrupt the first layer bonding in order to stabilize multilayer adsorption, so the 
 surface does not form an epitaxial multilayer. On Pt(111) the situation is different 
 \cite{thurmer2008,thurmer2008b}. In this case adsorption of more water
produces 3D islands of hexagonal or cubic ice  
(Fig.  \ref{fig:stm_structure}(h)) , reconstructing the original water film \cite{zimbitas2005,zimbitas2006}. 
Reconstruction of the first layer is driven by the increased H bonding to the multilayer,
but no detailed models are available for this buried interface, although a similar mechanism 
also operates on Ni(111) \cite{gallagher2007}. 
Removing uncoordinated OH groups in the first layer by locking them up in a tightly H bonded H$_2$O-OH 
structure prevents the first layer relaxing 
and instead clusters are formed \cite{zimbitas2008}.

\section{Discussion and Outlook: Questions we can and cannot answer}

The last few years has seen tremendous progress in basic understanding of water at surfaces, 
as the examples discussed illustrate. 
Although well-defined surface science style studies can be painstaking they are nonetheless 
incredibly useful as they reveal precisely at the molecular level how water behaves at interfaces. 
This is highly valuable information and some of the key questions these studies now allow 
us to answer, which just a few years ago we could only speculate upon, are:

\emph{(i) Where do water molecules preferentially adsorb on metal surfaces?} 
As we have seen, STM reveals that isolated water monomers 
adsorb above individual metal atoms of the substrate. 
This is consistent with DFT studies on many close-packed metal surfaces. 
DFT further predicts that in the most stable orientation the dipole moment of the water 
molecule is almost parallel to the surface (Fig.  \ref{fig:clusters}(c)), so 
as to facilitate interaction of the lone pair 1b$_1$ orbital of water with the substrate.
This basic adsorption structure for the water monomer is a key structural element 
of all water-metal overlayer adsorption
models characterised to date. 
However, perfect 2D H bonded networks cannot be built exclusively from flat-lying water
monomers because this would lead to too many O-H bonds being parallel to
the surface. Thus all extended overlayer structures characterised to date have a combination
of flat-lying water monomers at atop sites with other water molecules with their dipoles aligned more
or less with the surface normal.

\emph{(ii) How strong are interfacial H bonds?} The strength and length of H bonds
within metal-adsorbed clusters and overlayers can vary greatly, being either stronger
or weaker than H bonds between water molecules in the gas phase. 
On the one hand, cooperative forces can lead to stronger H bonds at the interface 
with metals. For example, the H bond length in the water dimer 
is generally shorter when it is adsorbed on a metal surface  (Fig. \ref{fig:clusters}(f))
than it is in the gas phase. 
This is because the dipole moment of the water molecule donating the H bond
is enhanced through polarisation by the substrate. However, on the other
hand, interaction with the surface can weaken the H bonding within the overlayer
as a result of  ``orbital competition'' \cite{michaelides2007,michaelides2007b}.
The alternating O-O bond lengths within
the adsorbed water hexamer provide an example of this effect: water molecules 
that interact strongly with the substrate make relatively poor H bond acceptors
because the water lone pair orbital (1b$_1$) is involved both in bonding
with the surface and in accepting H bonds.

\emph{(iii) How important is the number of H bonds to the stability of an overlayer?} 
When it comes to
water on metals it's quality and not quantity of H bonds that matters: the most stable
structure is not necessarily the one with the most H bonds. 
This is seen clearly in the STM experiments of water hexamers. The
experimentally observed structure is the cyclic water hexamer (Fig. \ref{fig:clusters}(g)), yet
this hexamer has only 6 H bonds, 2 fewer than the so called cage isomer, 
which is the most stable isomer in the gas phase. 
Another example is the OH-H$_2$O overlayer on Cu(110) that we have just discussed. 
In this system the H bond each OH molecule could in principle $donate$ to an 
adjacent water molecule has been sacrificed so as to facilitate the reorientation needed to
strengthen the H bonds the OH $accepts$ from neighbouring water molecules. 
The implication of this observation is that the adsorbed OH molecule is a
much better acceptor of H bonds than it is a donor, a conclusion that likely applies to
adsorbed OH groups in 
general \cite{michaelides2001,karlberg2003}.

\emph{(iv) Does a bilayer form when water adsorbs on metals?} 
For the best part of the 1980s and 1990s the bilayer model was essentially the ``standard
model'' for water adsorption on metals. 
However, from the discussion so far it should be clear that when wetting layer 
structures are examined in detail there is scant evidence for the formation
of adsorbed bilayers. 
The $\sqrt3$ diffraction pattern observed on many metals turn out not to be the result of bilayers and indeed are 
often associated with partially dissociated OH-H$_2$O overlayers instead \cite{hodgson2009}.
The one system for which there is the most compelling evidence so far is
water on an alloy of platinum and tin  \cite{mcbride2011};
a surface with $\sqrt3$ corrugation that
was specifically tailored to facilitate bilayer adsorption. 
Despite the widespread assumption that water adsorbs in a bilayer structure, 
the fact that it doesn't and that there is no general model for overlayer 
adsorption should not really come as a surprise given the immensely
rich phase diagram of water. 

Beyond the questions we currently have credible answers to there are many more which we can't 
yet answer with confidence. Some particular questions that investigators have been thinking about for many years
but are now increasingly turning their attention to are: 

\emph{(i) What are the main factors governing the structure of water overlayers at metal surfaces?}
Obviously this is H bonding and water-metal bonding, and for
specific systems we can understand how this balance plays out as discussed.
However, it would be premature to conclude that we fully understand 
this balance without, for example, the capacity to correctly \emph{predict} what 
the wetting layer structure would be for water 
on some arbitrary unexamined surface. 
What structures, for example, would 2/3 ML of water form on Rh(110) 
or 0.5 ML of water on Pd(100)?
Much more well-defined work combining experiment and theory is needed before the
underlying physical principles controlling the structure of water at metals 
are understood well enough to make confident intuitive predictions about wetting
layer structures. 

\emph{(ii) What happens beyond the first layer?}  The discussion thus far has largely 
focussed on the first contact layer between water and metals (at cryogenic temperatures), 
where there remains much to learn. 
However, increasingly there is interest in understanding multilayer water
adsorption and ideally the structure and dynamics of liquid water films at ambient 
temperatures. 
Traditionally the study of water multilayers and liquid water structure has been 
challenging for both experiment and theory alike.
This is still the case, however, recent developments are encouraging. 
For example, the emergence of high pressure x-ray photoelectron spectroscopy
(e.g. Ref. \cite{yamamoto2008}), improvements in atomic force microscopy 
experiments and analysis for solid-liquid 
interfaces (e.g. Refs. \cite{voitchovsky2010,fukuma2010,watkins2011}), 
and faster computers and 
computer codes all bring atomistic understanding of solid-liquid interfaces a step 
closer \cite{liu2008,schnur2009,ikeshoji2011,liu2011}.
Although much work remains to be done, one of the key concepts that has emerged
already is that nucleation of a second layer of water on top of the contact 
layer is facilitated by the presence of OH groups of water molecules that either stick out of the contact layer away from
the surface (so called dangling OH bonds) or can easily reconstruct to do so in the presence of more water.
It will be interesting to see what other concepts emerge over the coming years.

\emph{(iii) How can we do a better job with theory?} 
It should be clear by now that in this area there is an almost symbiotic relationship
between experiment and theory, with e.g. STM often relying on DFT to come up 
with plausible low energy structural models. 
It's only appropriate then to comment on some of the challenges remaining for theory. 
Generally with computer simulations one wants to tackle larger systems, over longer timescales, and
with increased accuracy. 
Advances in algorithmic efficiency and computational power are consistently making simulations with 
larger system sizes and over long timescales possible, as illustrated by 
the recent DFT-based molecular dynamics simulations of solid-liquid interfaces
\cite{liu2008,sharma2008,cicero2008,liu2008b,schnur2009,ikeshoji2011,liu2011}.
Increased efficiency and computational power also makes it 
much easier to explore configurational space when attempting to characterise surface structures. 
Indeed ``blind'' or ``random'' structure searching, which has proved to be a very valuable 
approach in materials and pharmaceutical science (see e.g. Ref. \cite{pickard2011}), 
has great potential to speed up the characterisation of novel surface structures in general
and water-ice structures in particular. 
In terms of increased accuracy, DFT has been remarkably successful in working out 
the structures of water on metals and more often than not the most stable adsorption structure identified with DFT 
has matched experiment. 
However, the standard exchange-correlation functionals used in most routine water adsorption studies (usually 
generalised-gradient approximation functionals such as PBE and PW91) have various well-known limitations.
One of the key limitations is that van der Waals dispersion forces are not accounted for and the lack of dispersion 
has been a topic of much discussion in this area over the years (see e.g. Ref. \cite{carrasco2011}). 
It is encouraging that dispersion-based DFT approaches are now beginning to be applied to water-metal
adsorption studies \cite{hamada2010,carrasco2011,lew2011,tonigold2012} and also that
electronic structure approaches beyond DFT are emerging for water adsorption, such 
as the Random Phase Approximation (RPA) and quantum Monte Carlo (QMC) \cite{ma2011}. 
Another important issue is the role of quantum nuclear effects (zero point motion, quantum tunnelling, and delocalisation).
At present the vast majority of theoretical studies treat the nuclei as
classical-point-like particles and only electrons are treated quantum mechanically. 
However, there is growing evidence from experiment and theory that
the quantum nature of the proton --- particularly in H bonded configurations such as for 
water overlayers at surfaces --- can have a substantial effect on the structure and 
dynamics of H bonds (see e.g. \cite{kumagai2009,li2010,li2011,kumagai2012}).

To conclude, the last few years has seen tremendous progress in fundamental understanding
of water at interfaces, with well-defined studies at metal surfaces providing key insight at
the molecular level. 
Important concepts in relation to the structure of water at surfaces have emerged but there remains 
much to be understood in terms of further developing the basic physical principles that control 
the structure of water at interfaces and in using these principles to improve technological 
processes such as ice nucleation inhibition or water desalination.
Being able to routinely predict the structure of a water adlayer on a given (metal) surface is certainly 
something we should soon be able to do and thus we should then be in a position
to use this information to identify (or develop) substrates which could be used to
e.g. inhibit or accelerate ice nucleation or facilitate very rapid water diffusion for 
the purposes of desalination.

\section{Acknowledgments}

J. C. is a Ram\'on y Cajal fellow and Newton Alumnus supported by the Spanish Government
and The Royal Society, respectively. A. M. is supported by the European Research Council. 

% --- Bibliography

% -- Nature style: nature.bst
\bibliographystyle{nature}
% --

\bibliography{ice}

\end{document}